\documentstyle[11pt,newpasp,epsf]{article}
\def\edcomment#1{\iffalse\marginpar{\raggedright\sl#1\/}\else\relax\fi}
\reversemarginpar

\begin{document}
\title{A Sociological Study of the Optically Emitting Isolated Neutron 
Stars}
\author{Patrizia A. Caraveo}
\affil{Istituto di Fisica Cosmica "G. Occhialini", Via Bassini, 15, \\20133 
Milano, ITALY - pat@ifctr.mi.cnr.it\\
and Istituto Astronomico, Via Lancisi 29 00161 Roma, ITALY}

\begin{abstract}

Although less than 1 \% of all radio pulsars are detected at optical 
wavelengths, their optical emission can yield a wealth of information 
that is either very 
difficult or plainly impossible to obtain at other wavelengths.
\end{abstract}

\section{Introduction}

The sample of the optically emitting Isolated Neutron Stars (INS) is not a 
rapidly growing one.
In spite of non negligible observational efforts, no new objects 
have been positively detected in the last few years.  The last 
{\it new} identification dates back to 1997 when HST resolved the 
counterpart of PSR1055-52 (Mignani et al., 1997).\\
Studying the optical behaviour of INSs is certainly a challenging 
task. There are neither routine nor serendipitous discovery such 
as in radio and, at least up to a point, in X-rays, where a numerous 
community has ample access to observing facilities.\\
At variance with radio and X-ray wavelengths, in the optical 
domain there are no instruments dedicated to the study of INS. 
Moreover, the tiny kernel left by a SN explosion is, by and large, not perceived 
as a potentially interesting object by the community of the 
optical astronomers. 

\section{INS Sociology}

All astronomical objects score
differently along the electromagnetic spectrum, but INSs
seem to be a rather extreme case.  Very prominent in the radio domain,
they undergo a minimum in the optical while the rise again
in X-rays, only to reach another maximum in high-energy gamma-rays.\\
Neutron stars are not glamorous optical emitters: they tend to be 
faint, point-like sources 
with flat spectra (when at all measured) and no lines.
But these balls of iron, ideal for studying physics under extreme conditions, 
have fostered more Nobel prizes than any other celestial object. However,
with their
ultrathin atmosphere, they are hardly considered stars any more.
Lacking prominent lines, they defy
the traditional tools of the optical trade and would
rather require ad hoc, unconventional approaches to unveil  
their peculiarities such as, e.g. the behaviour of elongated atoms. 
Moreover, their study 
calls for the most powerful optical telescopes, whose observing 
time is very much in demand for other hot topics in 
astronomy. Thus, Isolated Neutron Stars get, at best, few percent of 
the precious observing time of a big optical telescope, to be 
compared with a significant fraction of the observing time (if not 
the totality) of a radio one and a 10-20 \% in X-rays. No wonder 
the family of X-ray emitting neutron stars is more numerous than that of the 
optical ones. (see e.g. Becker and Tr\"umper, 1997)\\
Table 1 presents a summary of the data 
available on INSs with an 
optical identification either secured on the basis of timing or proper motion (PM)
or proposed on the basis of positional coincidence (Pos).
The grand total remains 9 and recent efforts on PSR 1706-44  (Mignani et al, 1999)
and on the newly discovered 16 msec pulsar, PSR 0537-6910 (Mignani et al, 2000),
have not yet yielded positive identifications.
Quite a lot of work went also into the timing of PSR0656+14 and Geminga (Shearer et al
1997,1998) but the low S/N of the ground based data has severely hampered the
statistical significance of the results gathered so far.

\begin{table}
\caption{Summary of optical observations available for all the INSs identified so far}
\begin{center}
\begin{tabular}{|l|l|l|l|l|l|} \hline
{\em PULSAR}  & {\em V mag} & {\em Timing} & {\em Pol} & {\em Photometry} & ID  \\ \hline
CRAB         & 16.6  &  Ground, HST   & Y & J to U,spectrum & Timimg         \\	
PSR0540-69    &22.4 &  Ground, HST   & Y & I,R,V,B,U,spectrum & Timimg                 \\
PSR1509-58    &22.0&  negative    &  (1) & R,V,B & Pos                                  \\
VELA           &23.6 &  Ground      &  -      & R,V,B,U & Timing                      \\
GEMINGA        &25.5& Ground? (2)      &  - & I,R,V,B,spectrum(3)    & PM           \\
               &  &            &    & 555,432,342,190 &    \\
PSR0656+14     &25.1&  Ground? (2)    & - & I,R,V,B,   & Pos                  \\
               & &  .            &    & 555,130L  & PM(4) \\           
PSR1055-52    &24.9  &              & - & {\it 342} (5)    & Pos       \\
PSR0950+08    &27.1  &              & - & {\it 130L} (5)   & Pos       \\
PSR1929+10     &25.7&              & - &130L,{\it 342} (5)   & Pos     \\ \hline
\end{tabular}
\end{center}
(1) see Wagner, these Proceedings \\
(2) tentative results to be confirmed (Shearer et al., 1997,1998)\\
(3) very low S/N \\
(4) tentative result (Mignani et al, 1997) to be confirmed through HST\\
(5) the magnitude of these objects
refers to the filter in italic 
\end {table}

The optical behaviour of INSs is composite, ranging from non-thermal
emission for the younger objects (Crab, PSR 0540-69, Vela and , possibly, 
PSR 1509-58), to mostly thermal, for the remaining, older, ones.
Caraveo (1998) and Mignani (1998) have comprehensively reviewed the subject.
\\Now we want to tackle the problem from a different point of view : sociology 
vs. physics and astronomy.\\  
Table 1 shows that 
the optically identified NSs are few and generally faint.
Are these unfavourable characteristics enough to explain
the very limited interest (or lack thereof) enjoyed by INS in the 
optical domain?  
\subsection{Does the appeal of a class of celestial sources depend
upon their number?}  
High energy gamma-ray astronomy offers an interesting example. 
While the bona fide NSs detected as gamma-ray sources 
are seven (Thompson et al, 1997), the 
high energy Astronomy community considers neutron stars 
amongst the most interesting objects in the sky and finds it perfectly sound to 
devolve quite a lot of observating time (and effort) to their quest.
\\A different example, in the optical domain, could be that of gravitational lensing systems,
which attracted, quite correcly, an enormous interest when numbering in
the few. The same is true also for the MACHO events.

\subsection{...or their brightness?}
The sky is full of faint targets which 
get their share of astronomical attention, irrespective of their
consistency as a class. In fact, ever since Galileo, in astronomy
the faintest objects, the ones at the limits of every telescope,
are always the newest and most exciting.\\
Optically identified INSs are comparable both in number and in brighness
to the optically identified GRBs, to mention a very recent hot 
topic. However, GRBs do have lines and thus the classical astronomical tools
can be immediately applied to them, making it worthwhile to obtain
spectra of 26 mv objects.\\ An inspection to Table 1 shows that faintness
is not even a limitation for neutron stars. The best 
studied neutron star is certainly Geminga, which is also one of the 
very faintest, while comparatively little has been done on the Crab, by far 
the brightest NS, and the only accessible also to small telescopes.

\section{The brightest and the dimmest}
Here we shall use Crab and Geminga as test cases to show that 
the understanding of a NS behaviour depends more on the interest it arises
in the community than on anything else.

\subsection{Crab}
Soon after the discovery of the pulsating star, 
a rough evaluation of its spectrum, 
proper motion and secular decrease were announced (see Nasuti et al, 1996 for a complete list of 
the references).
The proper motion was eventually nailed down in 1977 (Wyckoff and Murray, 1977) 
but nothing
was done to obtain a decent spectrum of an easy target nor to better assess its 
secular decrease.\\    
While the Crab was extensively, and wrongly, used 
as a textbook example of the multiwavelength behaviour
of INSs, two signatures of the emitting mechanism(s), namely the pulsar spectrum
and the decrease of its total brightness, were
totally neglected, in spite of their potential interest
for the understanding of the physics of the
Crab pulsar.
A reasonable spectrum was eventually obtained by Nasuti et al.(1996),
showing a real flat power low continuum. Does this measurement 
exhaust the interest in the spectral behaviour of the Crab?\\ 
Crab is certainly the only NS  bright enough to allow
a higher resolution spectral study, with the aim of looking  for something
unique to the special physics of the pulsar and its surroundings.\\
On the contrary, obtaining high resolution pictures of the Crab Nebula 
and its pulsar is a rewarding exercise. This lead to a series of HST pictures
which have been recently used also to measure anew the pulsar proper motion
(Caraveo and Mignani, 1999), suggesting a possible link between
the pulsar proper motion and the X-ray jet structure.
While the study of the interaction of the pulsar with the nearby medium
has been extensively carried out, the precise photometry
of the pulsar was never pursued. This has left the pulsar secular decrease
as an open issue of the physiscs of the Crab 
(see Nasuti et al, 1996 for a complete discussion).

\subsection{Geminga}

The long chase for Geminga has been reviewed
by Bignami and Caraveo (1996 and references therein) from its discovery,
back in 1973 by NASA's SAS II, to the HST era. The milestones in the Geminga chase
have been\\
-1981: confirmation and positioning by COS-B\\
-1983: discovery of a possible X -ray counterpart by the Einstein Observatory\\
-1987: pinpointing of its possible optical counterpart, dubbed G"\\
-1992: discovery of the 237 msec periodicity in X-rays, followed by
similar results in the $\gamma$-ray domain, thus linking the X and $\gamma$-ray
sources (ROSAT and EGRET) \\
-1993: measurement of the proper motion of G", thus proving its INS origin \\
-1996: improvement of the $\gamma$-ray light curve when taking 
into account the proper motion (Mattox et al.1996), 
thus linking G" to the source of $\gamma$-rays.\\
After the HST first refourbishement mission, in 1993, Geminga has been 
extensively imaged: 
first, to measure the source
parallactic displacement (Caraveo et al. 1996),
then to collect multiband photometry data.
These HST observations, 
confirming and refining difficult measurements with ground-based 
instruments, have resulted in the spectral distribution of Fig.1(Mignani et al, 1998).
A broad feature, centered at 
$\lambda=5998 \AA$ and with a width of 1,300 $\AA$, appears superimposed 
to the Rayleigh-Jeans continuum, as extrapolated from the soft X-rays. 

\begin{figure}
\epsfysize= 7cm
\epsfbox{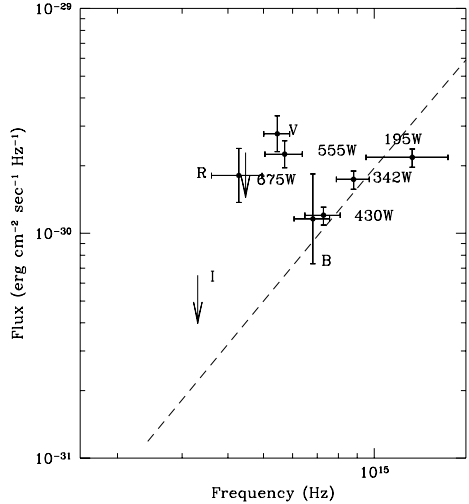}
%\centerline{\hbox{\psfig{figure=spectr-gem.ps,height=8cm,clip=}}}
\caption{Multiband photometry of Geminga. Three digits identify WFPC2/FOC
imaging filters. The dashed line represents the optical extrapolation of
the blackbody best-fitting the ROSAT data. For the best fitting temperature
of $5.77~10^{5} K$, and the measured distance of 157 pc, the emitting radius
is $R=10 km $.}
\end{figure}
If interpreted as an ion-cyclotron 
emission, this implies, for a pure H atmosphere a B field of $3.8~10^{11} G$  
(or $7.6~10^{11} G$ in the case of He, see Jacchia et al, 1999) 
not too far from to the value of $1.5~10^{12}$ obtained, theoretically, using 
Geminga's period and period derivative. \\This is the first time that the 
magnetic field of an INS is directly measured.\\
Moreover, the phenomenology of the source at high energies has been 
considerably enriched, owing to the very precise positioning of the 
optical counterpart.
The possibility to link HST data to the Hipparcos reference frame 
yielded the position of Geminga to an accuracy of 0.040 arcsec, a 
value unheard of for the optical position of a pulsar, or of any object 
this faint (Caraveo et al, 1998). This positional accuracy has allowed 
to phase together data collected over more than 20 years by SAS-2, 
COS-B and EGRET (Mattox et al, 1998).
The many "firsts" of Geminga have been summarized by Bignami (1998).\\
{\it Quite surprisingly, some of the key parameters of Geminga are now 
known with an accuracy better than that available for the Crab pulsar.} 
This is due in part to the remarkable stability of this object, 
which rendered possible to phase
together such a long time span of $\gamma$-ray data, in part to
the continous attention this object has been receiving by the 
astronomical community at large.

\subsection {Geminga-like sources}
The identification
of Geminga as a radio quiet INS broadened considerably
the framework of the multiwavelengths study of NSs, establishing 
a promising yet elusive template: the Geminga-like objects.
A Geminga-like source should be 
bright in $\gamma$ ray, conspicous in X-rays,
faint in the obtical, nul in radio. How many Gemingas are hidden in the
third EGRET catalogue (Hartman et al., 1999)? The only way to tell is to start a chase on
promising sources, possibly at middle galactic latitudes, to avoid 
far away objects, preferably in non crowded regions.
Imaging X-ray instruments, good resolution and high troughput are mandatory
to play the game with a reasonable efficiency.
The ESA XMM telescope, with its EPIC cameras, could play a significant role on this, as Einstein 
did for Geminga, 20 years ago.
\\
Caraveo, Bignami and Tr\"umper (1997) have further elaborated on this idea
applying the template to unidentified X-ray sources and recognizing
a handful of radio quiet INS candidates.\\
Once again the numbers are small, but they are going to grow rapidly.
On XMM, EPIC will certainly provide plenty
of serendipitous sources awaiting an identification.

\section{Making Neutron Stars Optically Appealing}

So far, the studies of the optical behaviour of NS have been
carried out mostly by "amateur" optical astronomers.
Indeed, those who develop a taste for the optical 
observations of Isolated Neutron Stars usually stumble into optical 
astronomy as a part of their multiwavelength approach aimed at 
the understanding of these fascinating objects.\\
Will it be possible to render the topic more appealing to the 
optical community, showing that NS are not just curious objects?\\
Will the optical Time Allocation Committees become more generous and invest observing 
time on the subject?\\
Will the community be able (or willing) to develop the needed tools?\\
Instruments devoted to timing are, of course, important, but 
timing, is just but one aspect of a multi-facet problem.
We have to apply all the tools of classical optical astronomy 
and possibly develop new ones.
The case of Geminga shows that endurance pays  but the process needs
to be accelerated.

\end{document}